\documentclass[aps,prl,reprint]{revtex4-1}
\usepackage{blindtext}

\usepackage{graphicx,epsfig,fullpage}
\usepackage{epstopdf}
\usepackage{subfigure}
\usepackage{wrapfig,color,enumitem}
\usepackage{amsmath,amssymb,latexsym,float}
\usepackage{mathrsfs,varioref,wrapfig,threeparttable}
\usepackage{dcolumn}   
 \newcolumntype{d}{D{.}{.}{-1}}
\usepackage{textcomp}
\usepackage{epsfig}
\usepackage{float}
\usepackage{tikz}
\usepackage{indentfirst}
\usepackage{lineno}
\usepackage{hyperref}
\usepackage{setspace}
\begin{document}
\title{Excitation of electrostatic solitary waves and surface waves in ion beam neutralization process}
\author{Nakul Nuwal$^\#$}
\email{nuwal2@illinois.edu}
\author{Igor D. Kaganovich$^\dagger$}
\author{Deborah A. Levin$^\#$}
\affiliation{$^\#$University of Illinois Urbana-Champaign, IL, USA, $^\dagger$Princeton Plasma Physics Lab, NJ, USA}

\begin{abstract}
Unusually long electrostatic solitary waves (ESWs) are discovered in 2D and 3D Particle-in-Cell studies of the process of ion beam neutralization by electron emission from filaments. These ESWs are long because trapped and untrapped electron density perturbations  nearly compensate each other.  Surface waves were discovered in the process of neutralization  but were only observed in 3D simulations. This is because the phase velocity of surface waves in a 2D geometry is higher than in 3D giving  the high-energy electrons  generated  upstream near the electron source enough energy to excite such waves only in  3D cylindrical beams.
\end{abstract}
\maketitle
\setlength\parindent{24pt}
Ion beams are used in various  applications such as particle accelerators, ion-thrusters\cite{Wang2012,jambunathan2018chaos,nuwal2020kinetic}, nanopantography\cite{chen2019nearly}, and ion-implantation\cite{conrad1989plasma}. The space-charge in the beam has to be neutralized to generate a focused ion beam, which is best done by a background plasma\cite{dudin2009simultaneous,kaganovich2001nonlinear,berdanier2015intense}. However, the complexity incurred by the generation of background plasma inside of a high-vacuum accelerator can be avoided by using an external filament electron source, which is easy to install in experiments. The neutralization by a filament electron source, however, leads to the excitation of space-charge waves, such as electrostatic solitary waves (ESWs), which limit the beam's space-charge neutralization\cite{lan2020neutralization2}. While the ESWs have been studied theoretically for many decades\cite{ng2005bernstein,chen2003width,chen2004bernstein}, their behavior has gained recent interest since the discovery of their ubiquitous formation in the process of charge neutralization of ion beams \cite{lan2019electrostatic,lan2020neutralization2}. The ESWs, which we also observed in our numerical simulations, have  mostly been studied in 1D in the past\cite{bernstein1957exact,hutchinson2017electron,chen2003width}. In this letter, however, using our heterogeneous high-performance-computing capable PIC code CHAOS\cite{jambunathan2018chaos}, we studied the formation of these ESWs in 2D planar and 3D cylindrical beams. We show that the previously developed 1D theoretical models can describe the long ESWs that form in 2D planar beams. However, they cannot describe the oblong axisymmetric ESWs that form in 3D cylindrical beams (see Fig. 13 of Ref. \cite{nuwal2021PRE}). Further, by conducting a theoretical analysis of these modes similar to the original BGK treatment\cite{bernstein1957exact}, we show that because of a non-Maxwellian electron distribution function (EVDF) formed by neutralizing electrons in the beam, \emph{ the ESWs are much longer} than those observed in the previous theoretical \cite{chen2002bgk,goldman2007theory} and experimental works\cite{lefebvre2010laboratory}. 

\begin{figure}[!h]
\centering
\includegraphics[trim = 0.14cm 0.14cm 0.14cm 0.14cm, clip,width = 0.42\textwidth]{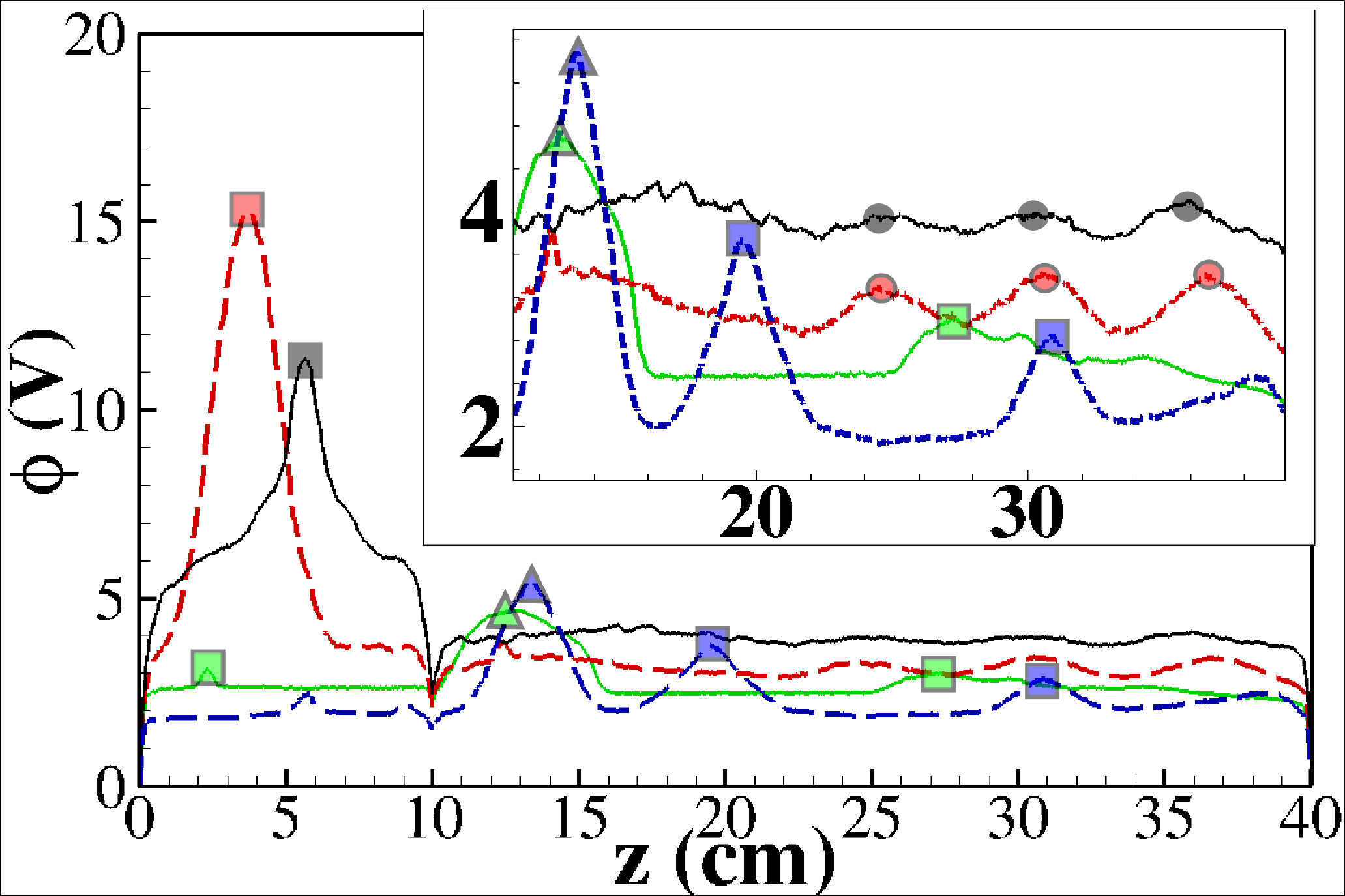}       
\caption{Electric potential profiles along the beam axis at $t = 4.72 \ \mu s$, when all cases have achieved at least $80\%$ of neutralization. Cases: 2D (blue), $a=2.5$ mm 3D (green), $a=5.0$ mm 3D (red), and $a=7.5$ mm 3D (black); $a$ is the beam radius. Wave features: ESWs (square boxes), electron-holes (triangles), and surface waves (circles). }
\label{fig:Phi_profiles}
\end{figure}

Additionally, in 3D simulations, we observed  the excitation  of the Trivelpiece-Gould (TG) surface waves\cite{trivelpiece1959space} in the process of space charge neutralization of an ion beam. The TG waves have been experimentally studied  for  plasma columns in waveguides \cite{lynov1979observations,moisan1982experimental} where they were excited by an externally applied field. In this work, however, an ion beam together with  neutralizing electrons form a plasma column that can serve as a plasma waveguide for surface waves \cite{moisan1982experimental}; i.e., emitted electrons from a filament form  an electron beam at early stages of neutralization process that is uniquely suited to spontaneously excite such surface waves with a high phase speed. At later time in the neutralization process, these high energy electrons leave the simulation's domain due to wall losses. This prevents the decay of TG waves due to the Landau damping by fast electrons, which allows TG waves to exist for very long times. Finally, we derived a dispersion relation for the surface waves in a 2D planar ion beam in a metal chamber and identified that the  phase speed of the surface waves in 2D planar ion beams is sufficiently different than 3D, explaining why surface waves are only observed in 3D.      

Particle-in-Cell (PIC) simulations were performed in a 2D domain of $ 6 \times 40$ $\rm cm^{3}$ and a 3D domain of $3 \times 3 \times 40$ $\rm cm^{3}$ (see Figs. 2(a) and 2(b) of Ref. \cite{nuwal2021PRE}), where a monoenergetic argon ion beam of 38 keV was injected  at the center of the $z = 0$ mm plane from a rectangular(circular) region for 2D(3D) case(s). A uniform grid with cell sizes ranging from $0.3\lambda_{\rm D}$ to $0.91\lambda_{\rm D}$ was used for both 2D and 3D cases, here, $\lambda_{\rm D}$ is the local Debye length at the end of neutralization process when the electron density is highest. The 40 cm length of our domain and about $2.5$ mm beam inlet radius (half-width for 2D) is used to be comparable to the previously conducted plasma neutralization  experiments\cite{stepanov2016dynamics}. Note that this study is also relevant to  nanopantography applications\cite{chen2019nearly}. The electrons with a $T_e = 2$~eV Maxwellian EVDF and current, $I_e = I_{\rm Ar^+}/3$, $I_{\rm Ar^+}$ being the ion beam current, were introduced at the beam axis at $z = 10$ cm. The neutralization process of the ion beam happens in two phases: (1) rapid neutralization, which occurs when the beam potential, $\phi >> k_BT_e/e$, and (2) slow neutralization when $\phi \approx k_BT_e/e$. After the initial rapid neutralization phase, the rate of neutralization slows down considerably because the remaining low unneutralized potential can no longer trap all emitted electrons and high-energy electrons escape to the walls\cite{lan2020neutralization}.    

More than $80\%$ of neutralization is achieved by $t = 4.0 \ \mu s$. Figure \ref{fig:Phi_profiles} shows the electric potential profiles at the end of the first phase in the neutralization process, in which ESWs form both upstream and downstream of the electron source for the 2D and 3D beam cases with $a = 2.5$ mm, whereas ESWs only form in the upstream and surface waves in the downstream regions of the electron source for 3D cases with $a = 5.0$ and 7.5 mm, as shown in the figure inset. 
Space-charge sheaths form at $z_{\rm min}$  and $z_{\rm max}$ boundaries and reflect most electrons and produce two electron streams formed inside the ion beam. The two-stream instability between  two electron streams results in the formation of ESWs, as was reported in Refs.\cite{lan2020neutralization2,omura1996electron,roberts1967nonlinear}. 
For the neutralization process, the ESWs were observed to be very robust: they are ubiquitously excited during the neutralization process, survive many collisions between themselves and reflection from the boundaries.\cite{lan2020neutralization2}

\begin{figure}
\centering
\subfigure{\label{fig:2D_soliton_schematic_t312}
        \includegraphics[trim = 0.14cm 0.1cm 0.14cm 0.14cm, clip,width = 0.48\textwidth]{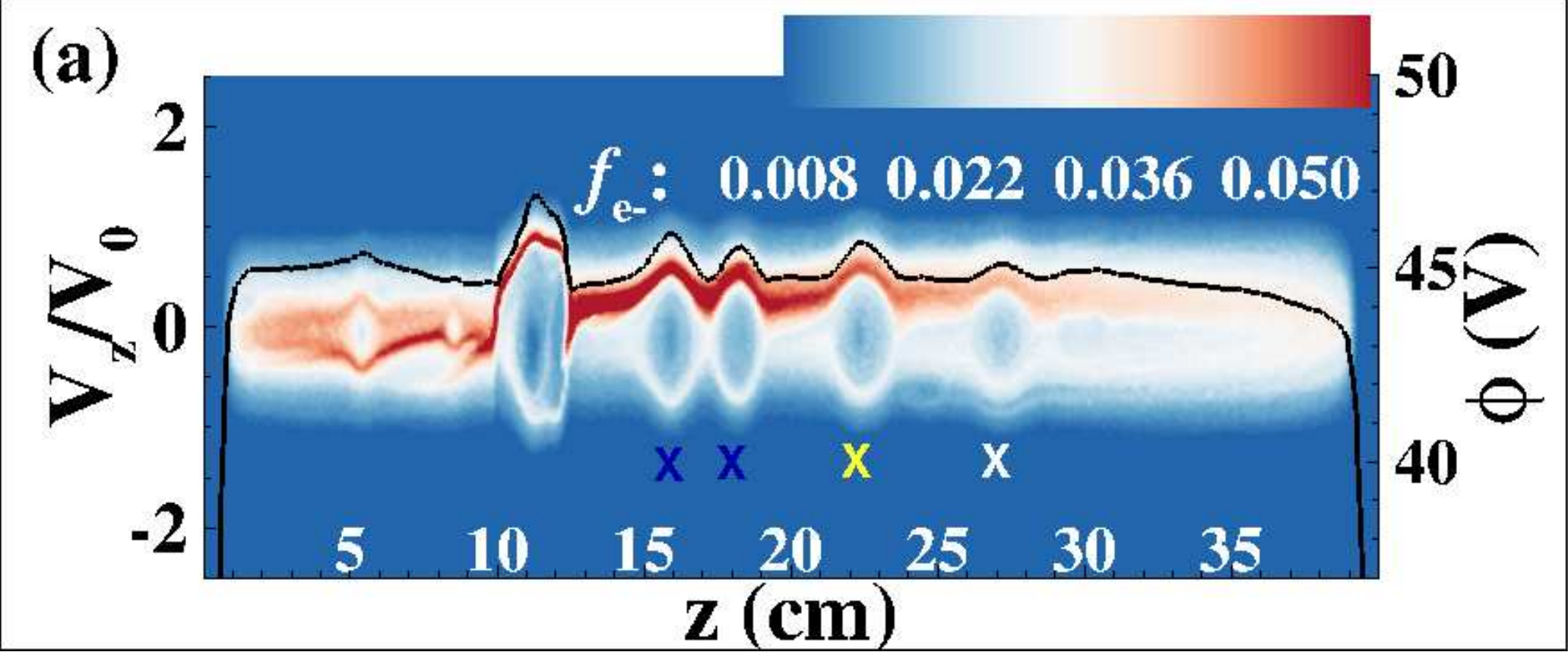}}
  \subfigure{\label{fig:2D_soliton_movement_timeevol}
        \includegraphics[trim = 0.2cm 0.14cm 0.2cm 0.3cm, clip,width = 0.48\textwidth]{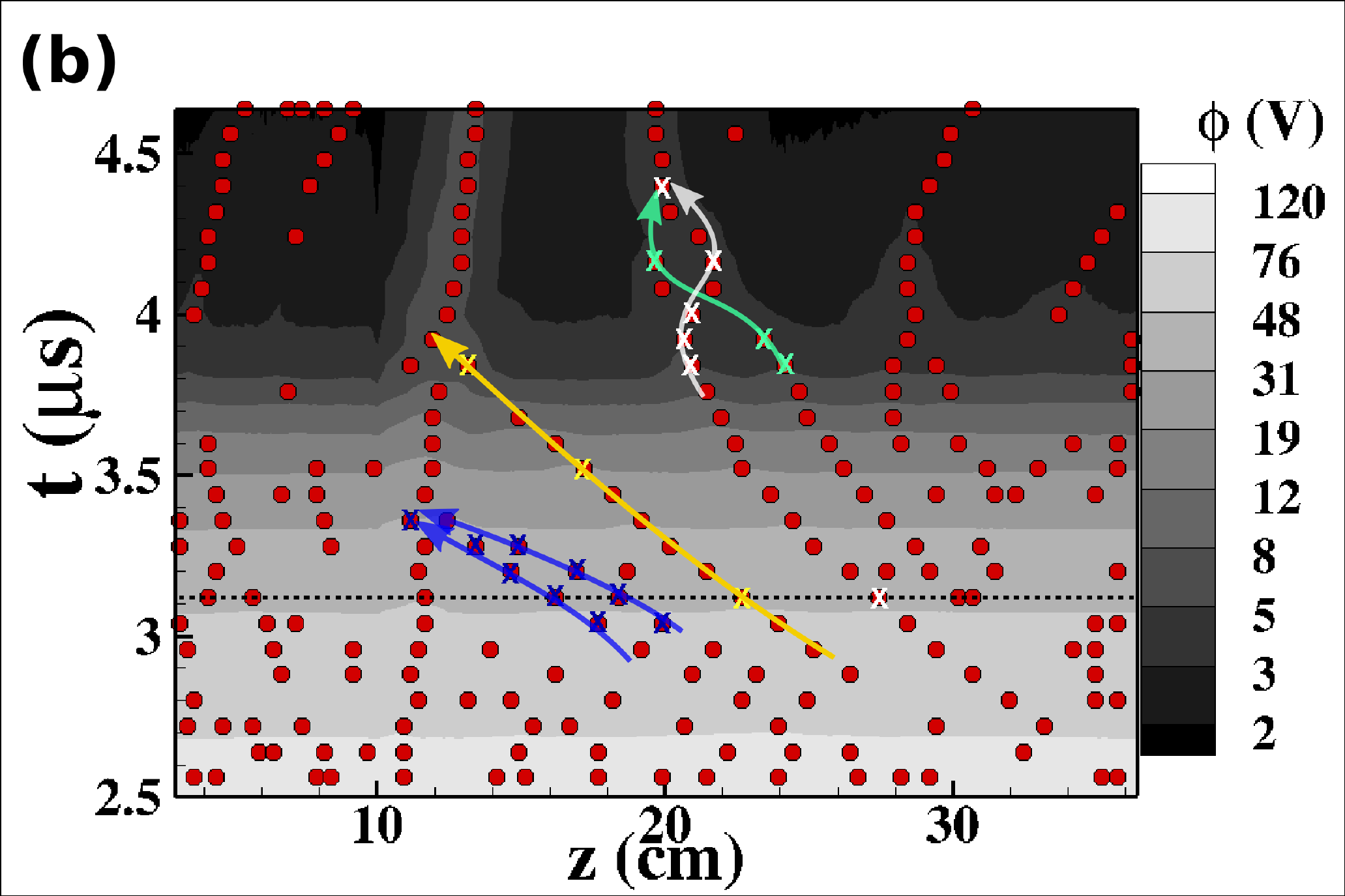}}
\caption{ (a) Electron distribution in phase space, $f_{e-}$ normalized by $n_0 = 1.75\times10^{14} \ {\rm m^{-3}}$, at $t = 3.12 \ {\rm \mu s}$. The black solid line in (a) shows electric potential on the right vertical-axis. (b) Time evolution of ESWs positions, shown by red circles, along the beam axis, $(x,y) = (0.18,3.0)$ cm, for the 2D planar beam case. In (b), ESWs shown by blue and yellow \textbf{`X'}s eventually merge with the large electron-hole near the electron source at $z = 10$ cm, and ESWs shown by white and green \textbf{`X'}s merge with each other.   }
\label{fig:2D_soliton_movement_timeevol_all}
\end{figure}

These ESWs can be easily recognized  as vortex-like structures in the electron phase space, as shown in Fig. \ref{fig:2D_soliton_schematic_t312}. 
These ESWs (their positions shown by the red-dots in Fig. \ref{fig:2D_soliton_movement_timeevol} for discrete number of times for 2D planar beam case) move, collide, and merge with each other to eventually form new ESWs that are of larger lengths than the original ones. Among the multiple ESWs in Fig. \ref{fig:2D_soliton_movement_timeevol}, the ESW shown by the yellow `\textbf{X}' is of particular interest because it moves with nearly a constant speed, and does not collide or merge with other ESWs for a long time. Because its electric potential amplitude was found to be similar to the other ESWs in the system, the yellow `\textbf{X}' ESW is representative of other ESWs in the 2D planar beam case and is further analyzed below. 

\begin{figure}
\centering
\subfigure{\label{fig:Soliton_2D}
        \includegraphics[trim = 0.10cm 0.1cm 0.10cm 0.05cm, clip,width = 0.48\textwidth]{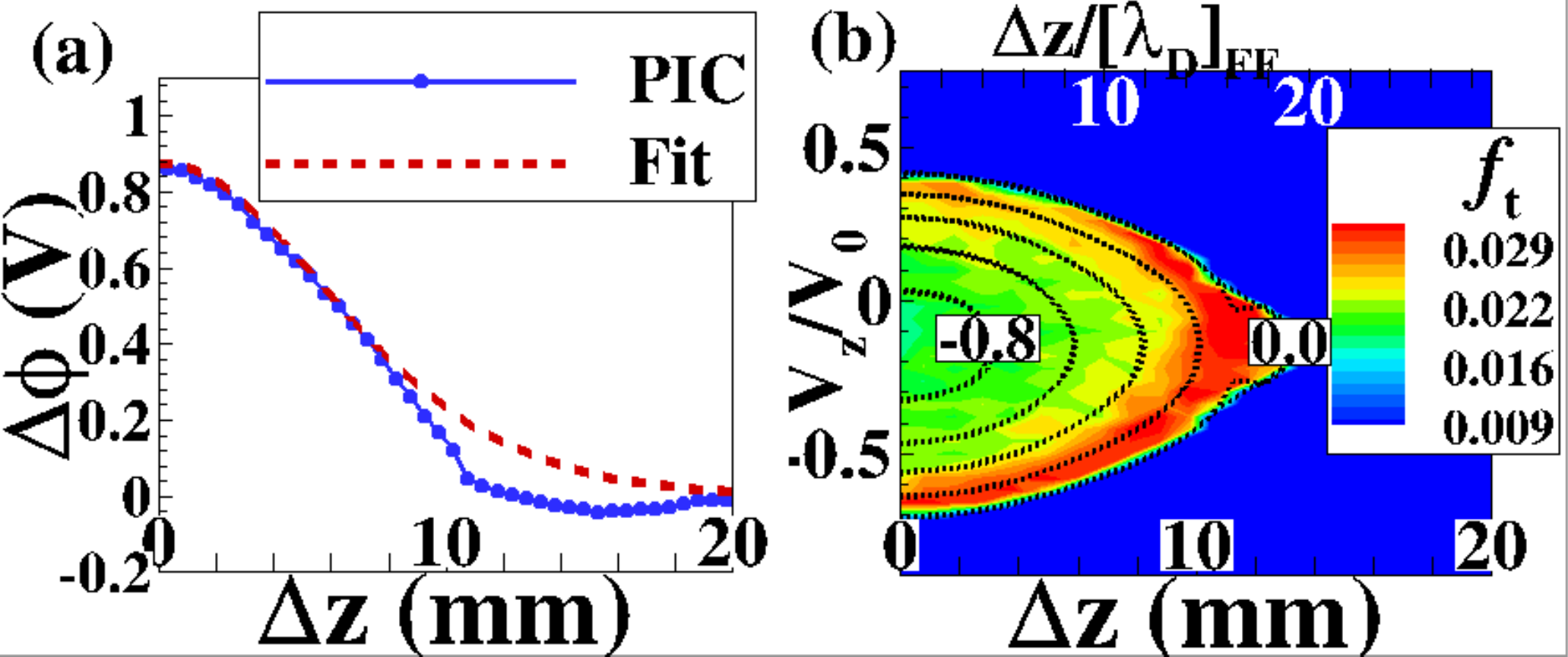}}
  \subfigure{\label{fig:BGK_2D}
        \includegraphics[trim = 0.14cm 0.10cm 0.14cm 0.10cm, clip,width = 0.48\textwidth]{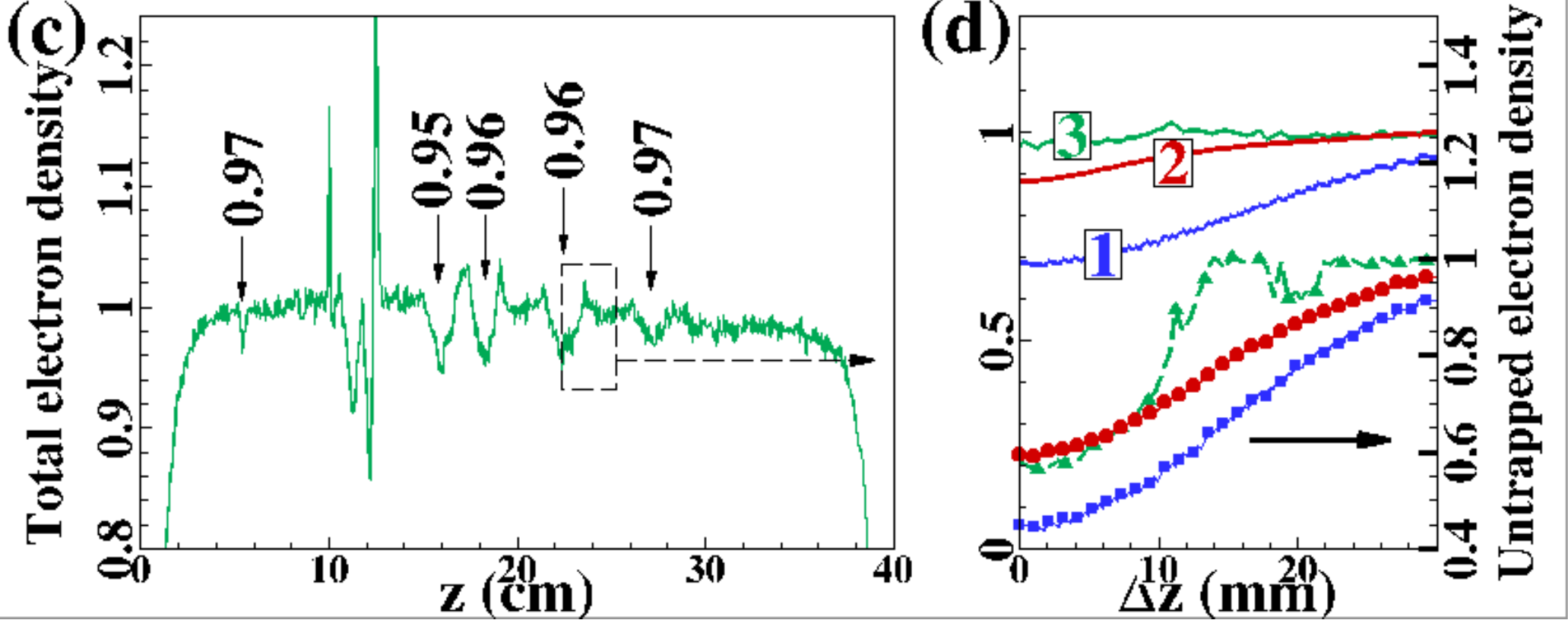}}
\caption{ In (a), the potential profile and the trapped electron distribution in phase space, $f_t$ (normalized by $n_0 = 1.75\times10^{14} \ {\rm m^{-3}}$),  are shown for ESW marked with yellow `\textbf{X}' in Fig. \ref{fig:2D_soliton_schematic_t312}. (b): Dashed black lines  show constant $E_{\rm tot}$ (eV) contour lines with increments of 0.2 eV. \textcolor{black}{(c) shows the electron density profile normalized by $1.5\times10^{14} \ {\rm m^{-3}}$.   (d) Comparison of predicted total (untrapped) electron density based on Eq. \ref{eq:untrapped_EVDF_selfsimilar} using a  1. Maxwellian (blue) and 2. non-Maxwellian (red) $f_{\rm FF}$ and 3. the PIC result (green). }  }
\label{fig:Soliton_structure_2D_3D}
\end{figure}

Because of their collisions and merging, ESW lengths in both 2D and 3D beams increase  to become between 30-50 local Debye lengths long, which is unusually long when compared to the other studies\cite{lefebvre2010laboratory,chen2002bgk,hutchinson2017electron}. 
Since the beam radius is small compared with the ESW length,  we analyze the yellow \textbf{`X'} ESW at $t = 3.12 \ \mu s$ in Fig. \ref{fig:Soliton_2D} using 1D BGK theory\cite{bernstein1957exact}, by fitting an analytical electric potential profile to the PIC result\cite{schamel1979theory}, 
\begin{align}\label{eq:phi_fit}
    \Delta \phi (\Delta z) = {\phi_0} {\rm sech^4}\left(\frac{b_s\Delta z}{\lambda_{\rm D_0}}\right),
\end{align}
where we obtain $\phi_0 = 0.876$ V and $b_s = 0.05$ as the fit parameters, $\lambda_{\rm D_0} = 0.78$ mm is the Debye length at $n_e = 1.75\times10^{14} \ {\rm m^{-3}}$ and $T_e = 2$ eV. 
The comparison of the fit with PIC is shown in Fig. \ref{fig:Soliton_2D}, where it can be seen that it compares well for $\Delta z < 10$ mm. In Eq.~\ref{eq:phi_fit}, $\Delta \phi = \phi(\Delta z) -  \phi_{\rm base}$, where $\phi_{\rm base}$ is the  potential outside the ESW at $\Delta z = 15$ mm. 
In Fig. 3(b), the contour lines of the constant total energy, $E_{\rm tot}$, are shown by dashed black lines and are superimposed on color-plot of the EVDF of the trapped-in-the wave electrons, $f_t$, in the phase space for the ESW marked by yellow `\textbf{X}' in Fig. 2. They indicate a small variation in electron populations at different $E_{\rm tot}$ values, which is qualitatively similar to the `shallow' ESWs discussed in Ref. \cite{hutchinson2017electron}. Given that the EVDF of electrons far from the ESW, $f_{\rm FF}$, is known, the untrapped EVDF reads,
\begin{align}\label{eq:untrapped_EVDF_selfsimilar}
    f_{ut}\left(v_z ,\Delta z\right) = f_{\rm FF}\left( \text{sgn}{(v_z)}  \sqrt{v_z^2-2\frac{e}{m_e}\Delta \phi(\Delta z)}\right) 
\end{align}
where $f_{ut} (v_z,\Delta z)$ is the EVDF of untrapped electrons at location $\Delta z$.
We can determine the electron density profiles of both the trapped and untrapped electrons as a function of potential using Eq. \ref{eq:untrapped_EVDF_selfsimilar} for untrapped and a similar equation for trapped electrons where the EVDF is taken from the PIC simulations results at the location of the potential maximum. 
We now compare the effect of different $f_{\rm FF}$ on ESW: 1. a Maxwellian EVDF with $T_e =1$ eV chosen to fit to the PIC result, and 2. non-Maxwellian EVDF taken directly from the PIC simulation results in Fig. 3(b) at $\Delta z = 15$ mm (also see Fig. 10(c) of Ref. \cite{nuwal2021PRE} for further details).  

The ESWs generated in our PIC simulations seem to have formed having a very small electron density perturbation of the order of 3-5\%, as shown by depletion in the normalized electron density in Fig. 3(c). \textcolor{black}{This small density perturbation in ESWs is allowed due to the non-Maxwellian background EVDF in the beam. This is demonstrated in Fig. 3(d) where the total electron density perturbation from Eq. \ref{eq:untrapped_EVDF_selfsimilar} applied to a non-Maxwellian $f_{\rm FF}$ (`2') is closer to the PIC result (`3') than a Maxwellian $f_{\rm FF}$ (`1'). This difference in the magnitude of these perturbations is entirely due to a lower electron density decline of untrapped electrons towards the center of the ESW for a non-Maxwellian $f_{\rm FF}$ than a Maxwellian one, as shown on the secondary axis of Fig. 3(d). Since the EVDF at the center and the electric potential profile are taken from the PIC result, trapped electron profiles are identical for both $f_{\rm FF}$ cases. \textcolor{black}{This small untrapped density decline nearly compensates the trapped density in the non-Maxwellian case, resulting in a very small total density perturbation.} This shows that a Maxwellian background EVDF would likely result in a larger electron density and electric potential perturbation that a non-Maxwellian one found in ion beams and this is the reason why such long ESWs with a small density and potential perturbations appear in our ion beam.}  
Present 1D BGK theory needs to be revised to be able to describe  ESWs in 3D  due to their highly complex electron trajectories (they are shown in Fig. 16 of Ref. \cite{nuwal2021PRE}).

\begin{figure}
\centering
\subfigure{\label{fig:5mm_case_waves}
        \includegraphics[trim = 0.16cm 0.14cm 0.16cm 0.15cm, clip,width = 0.40\textwidth]{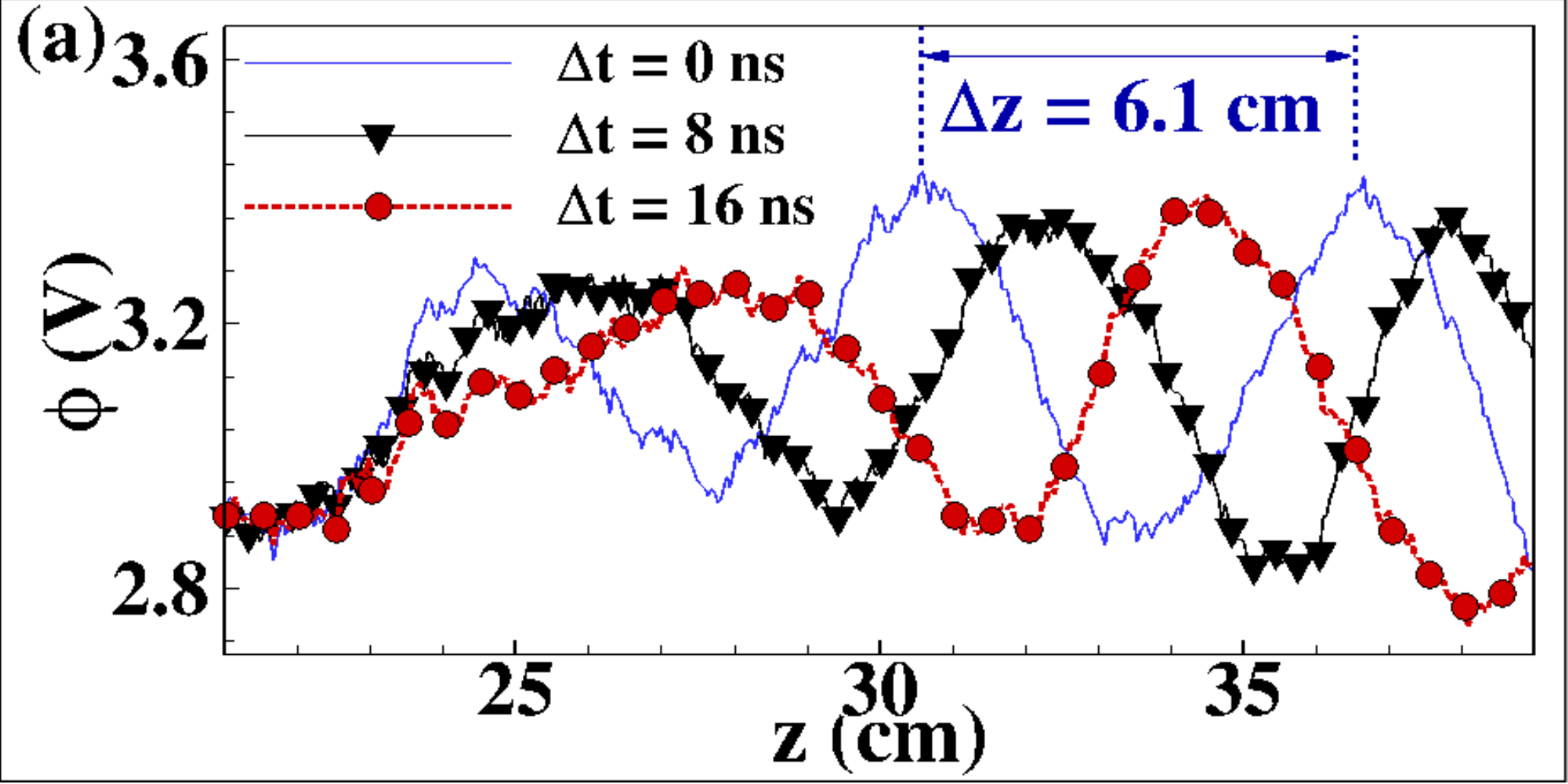}}
\subfigure{\label{fig:TG_phase_excitation}
        \includegraphics[trim = 0.14cm 0.1cm 0.1cm 0.14cm, clip,width = 0.40\textwidth]{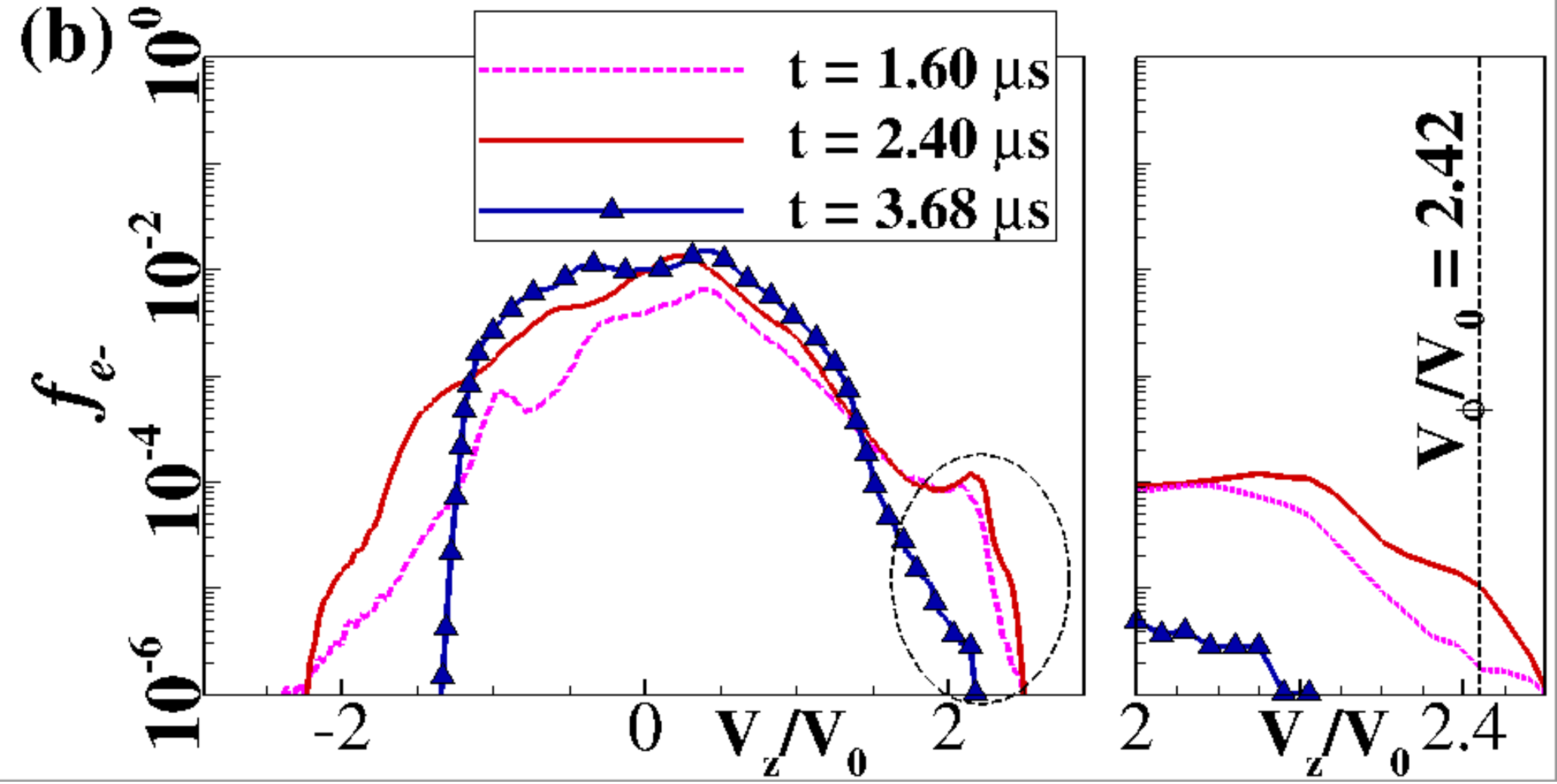}}      
\subfigure{\label{fig:TG_dispersion}
        \includegraphics[trim = 0.14cm 0.14cm 0.16cm 0.16cm, clip,width = 0.40\textwidth]{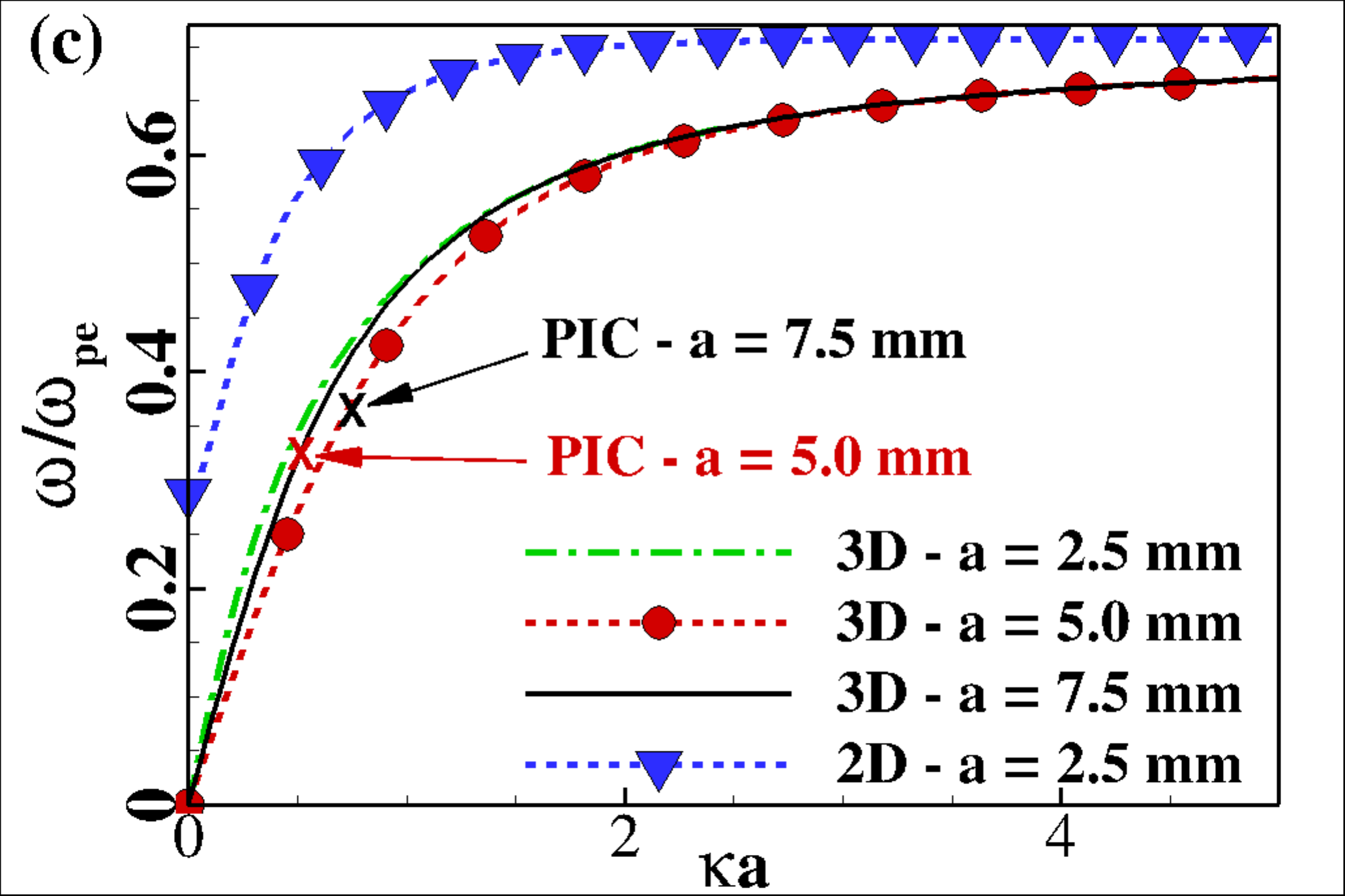}}  
\caption{(a) Time evolution of electric potential profile $\Delta t = t - 4.72 \ \mu s$. (b) Evolution of EVDF, $f_{e-}$ normalized by $n_0 = 1.75\times10^{14} \ {\rm m^{-3}}$,  for 3D beam case of $a = 5.0$ mm with a few electrons reaching  the surface wave phase velocity, $v_\phi = 2.42\times10^6$ m/s at $t = 2.40 \ \mu s$. (c) Dispersion curves for the surface waves in 3D cylindrical\cite{trivelpiece1959space} and 2D planar beams. }
\label{fig:TG_dispersion_wave}
\end{figure}

Turning to understanding the importance of 3D effects, ESWs were not found downstream of the electron source for the 3D simulations with beam radii $a = 5.0$ and $a=7.5$~mm. In these cases, the potential of the un-neutralized ion beam is higher than in the case with $a = 2.5$ mm, because there are more ions in beams of thicker radii. Correspondingly, in these cases, electrons gain velocities much higher than thinner beams. Our PIC simulations indicate that, as a result, the energetic electrons  excite high-phase-speed surface waves, as shown in Fig. \ref{fig:5mm_case_waves}. These waves have a long wavelength $\lambda_w = 6.1$ cm and high phase velocity, $v_\phi = 2.42 \times 10^6$ m/s, for the $a = 5.0$ mm case. Figure \ref{fig:TG_phase_excitation} shows the temporal evolution of the  EVDF  for the $a = 5.0$~mm case where a small number of electrons gain velocity close to $2.42 \times10^6$ m/s at about $t = 2.40 \ \mu s$ and excite the surface wave of that phase velocity. 

We compared the surface waves parameters obtained from the simulations and the theory using the dispersion relation for axisymmetric surface waves derived by Trivelpiece and Gould (TG)\cite{trivelpiece1959space}, as shown in Fig. \ref{fig:TG_dispersion}.  This shows that axisymmetric TG surface waves become excited for 3D cases only with $a = 5.0$ and $7.5$ mm. Although high-velocity electrons excite these waves, with the decrease in electric potential, the ion beam space charge's ability to generate high-energy electrons declines\cite{lan2020neutralization} with time. This leads to a decline in the electron temperature inside the beam, shown by the narrowing of EVDF with time in Fig. \ref{fig:TG_phase_excitation}, which leaves the previously excited TG waves in the domain with no high-velocity electrons in resonance with the waves that could quench them through Landau damping. \textcolor{black}{This also explains the absence of TG waves in the $a = 2.5$ mm 3D case where the ion beam space-charge generates electrons with high-enough velocities to dampen the low phase speed surface waves that may exist for such small radius beam(see Fig. 18(b) of Ref. \cite{nuwal2021PRE}).} For more details see Sec. 4 of Ref. \cite{nuwal2021PRE}.

Such surface waves, however, were not observed in the 2D planar beam\cite{lan2020neutralization}. Following a procedure similar to Vedenov\cite{vedenov1965solid}, Krall\cite{krall1973principles}, and Trivelpiece-Gould\cite{trivelpiece1959space}, we derived the dispersion relation of a 2D planar beam, 
\begin{align}\label{eq:2D_dispersion}
    \frac{\omega}{\omega_{pe}} = \frac{1}{\left[1+ \frac{\tanh{[(b-a)\kappa]}}{\tanh{a\kappa}}\right]^{1/2}},
\end{align}
where $a$ and $b$ are the half-widths of plasma beam and channel, respectively (See Appendix C of Ref. \cite{nuwal2021PRE} for details). Equation \ref{eq:2D_dispersion} converges to $1/\sqrt{2}$ for $a\to \infty$, $b\to \infty$, and $a << b$, as expected for a planar surface wave in a vacuum with no conducting walls\cite{vedenov1965solid}.  As seen from Fig. \ref{fig:TG_dispersion}, the surface waves in the 2D planar beam have a higher phase speed, $v_\phi = \omega/\kappa$, than the TG surface waves for cylindrical beams. This means that, when comparing 2D and 3D surface waves dispersion for $a = 2.5$ mm, electrons with about 2.5 times higher velocities than 3D beam would be required to excite these waves in a 2D beam. Since such high velocity electrons are not present when the electrons neutralize the ion beam in the 2D simulation, the surface waves were not observed in that case.  

Finally, in our PIC simulations, we were able to artificially excite these surface waves when we introduced a small amount of emitted electrons with velocities higher than the electron thermal velocity of $T_e = 2$ eV in  2D and 3D beam cases of $a = 2.5$ mm, where the surface waves were not spontaneously excited originally (see Fig. 21 of Ref. \cite{nuwal2021PRE}). This shows that a small amount of high velocity electrons may excite a surface wave and affirms that the TG waves in our other 3D cases were excited by high energy electrons.

\textcolor{black}{In summary, in our kinetic PIC modeling we observed a curious fact that ESWs with axial lengths of 30-50 Debye lengths can be generated during neutralization of an ion beam by electron emission. In a 2D geometry, such long ESWs can be approximately described by a modified 1D BGK theory because the ESW length is much larger than the beam width.  The large length of ESW is determined to be a consequence of a surprising near compensation of density variation along the ESW by trapped and untrapped electrons because of the non-Maxwellian background EVDF. 
We also observed the generation of TG surface waves in cylindrical ion beams during neutralization process simulations in 3D geometry; whereas these waves were not observed in 2D geometry. Analysis shows that this can be explained by the fact that surface waves have a much higher phase velocity in 2D than in 3D.  }

This research was supported by the US Department of Energy Award DE-SC00021348. I.D.K. was supported by US Department of Energy contract DE-AC02-09CH11466. 




\bibliography{sample}
\bibliographystyle{aiaa}
\end{document}